\documentclass[twocolumn,superscriptaddress,showpacs,preprintnumbers,amsmath,amssymb,prl]{revtex4}
\usepackage{graphicx}
\usepackage{epsfig}
\usepackage{dcolumn}
\usepackage{bm}
\usepackage{longtable}%
\usepackage{dcolumn}
\usepackage{colordvi}\usepackage[usenames]{color}
\def\etal{{\it et al.}}
\newcommand{\header}[1]{{\em#1.---}}
\newcommand{\comment}[1]{#1}

\def\CEA{D\'epartement de Physique Nucl\'eaire, IRFU, CEA, Universit\'e Paris-Saclay, 91191 Gif-sur-Yvette, France}
\def\RIKEN{RIKEN Nishina Center, Hirosawa 2-1, Wako, Saitama 351-0198, Japan}
\def\CEAii{D\'epartement d'\'electronique des D\'etecteurs et d'Informatique pour la Physique, IRFU, CEA, Universit\'e Paris-Saclay, 91191 Gif-sur-Yvette, France}
\def\CEAiii{D\'epartement des Acc\'el\'erateurs, de Cryog\'enie et de Magn\'etisme, IRFU, CEA, Universit\'e Paris-Saclay, 91191 Gif-sur-Yvette, France}
\def\CEAiv{D\'epartement d'Ing\'enierie des Syst\`emes, IRFU, CEA, Universit\'e Paris-Saclay, 91191 Gif-sur-Yvette, France}
\def\CNS{Center for Nuclear Study, University of Tokyo, Hongo 7-3-1, Bunkyo, Tokyo 113-0033, Japan}

\def\Padova{Dipartimento di Fisica e Astronomia ``G.~Galilei'' and INFN - Sezione di Padova, Via Marzolo 8, 35131 Padova, Italy}
\def\FAMN{Departamento de F\'{\i}sica At\'omica, Molecular y Nuclear, Facultad de F\'{\i}sica, Universidad de Sevilla, Apartado 1065, E-41080 Sevilla, Spain}
\def\TUDA{Department of Physics, Institut f\"ur Kernphysik, Technische Universit\"{a}t Darmstadt, D-64289 Darmstadt, Germany}
\def\CHINA{School of Physics, Peking University, Beijing 100871, China}
\def\ORSAY{Institut de Physique Nucl\'eaire, IN2P3-CNRS, Universit\'e Paris-Sud, Universit\'e Paris-Saclay, 91406 Orsay Cedex, France}
\def\LPC{LPC Caen, ENSICAEN, CNRS/IN2P3, Universit\'e de Caen, Normandie Universit\'e, 14050 Caen, France}
\def\TOHOKU{Department of Physics, Tohoku University, Miyagi 980-8578, Japan}
\def\Titech{Department of Physics, Tokyo Institute of Technology, 2-12-1 O-Okayama, Meguro, Tokyo 152-8551, Japan}
\def\KYOTO{Department of Physics, Kyoto University, Kyoto 606-8502, Japan}
\def\KYUSHU{Department of Physics, Kyushu University, Nishi, Fukuoka 819-0367, Japan}
\def\MIYAZAKI{Department of Applied Physics, University of Miyazaki, Gakuen-Kibanadai-Nishi 1-1, Miyazaki 889-2192, Japan}
\def\EHW{Department of Physics, Ehwa Womans University}
\def\TOKYO{Department of Physics, University of Tokyo, Hongo 7-3-1, Bunkyo, Tokyo 113-0033, Japan}
\def\TUM{Department of Physics, Technische Universit\"{a}t Munchen, Germany}
\def\FAMN{Departamento de F\'{\i}sica At\'omica, Molecular y Nuclear, Facultad de F\'{\i}sica, Universidad de Sevilla, Apartado 1065, E-41080 Sevilla, Spain}
\def\NSCL{FRIB, Michigan State University, East Lansing, Michigan 48824, USA}
\def\Hope{Physics Department, Hope College, Holland, Michigan 49423, USA}
\begin{document}
\title{{\boldmath{}Mass, spectroscopy and two-neutron decay of $^{16}$Be}}
\author{B.~Monteagudo}\affiliation{\LPC}\affiliation{\NSCL}\affiliation{\Hope\footnote{Present address}}
\author{F.M.~Marqu\'es}\affiliation{\LPC}
\author{J.~Gibelin}\affiliation{\LPC}
\author{N.A.~Orr}\affiliation{\LPC}
\author{A.~Corsi}\affiliation{\CEA}
\author{Y.~Kubota}\affiliation{\RIKEN}\affiliation{\CNS}\affiliation{\TUDA}
\author{J.~Casal}\affiliation{\Padova}\affiliation{\FAMN}
\author{J.~G\'omez-Camacho}\affiliation{\FAMN}
\author{G.~Authelet}\affiliation{\CEAiii}
\author{H.~Baba}\affiliation{\RIKEN}
\author{C.~Caesar}\affiliation{\TUDA}
\author{D.~Calvet}\affiliation{\CEAii}
\author{A.~Delbart}\affiliation{\CEAii}
\author{M.~Dozono}\affiliation{\CNS}
\author{J.~Feng}\affiliation{\CHINA}
\author{F.~Flavigny}\affiliation{\ORSAY}
\author{J.-M.~Gheller}\affiliation{\CEAiii}
\author{A.~Giganon}\affiliation{\CEAii}
\author{A.~Gillibert}\affiliation{\CEA}
\author{K.~Hasegawa}\affiliation{\TOHOKU}
\author{T.~Isobe}\affiliation{\RIKEN}
\author{Y.~Kanaya}\affiliation{\MIYAZAKI}
\author{S.~Kawakami}\affiliation{\MIYAZAKI}
\author{D.~Kim}\affiliation{\EHW}
\author{Y.~Kiyokawa}\affiliation{\CNS}
\author{M.~Kobayashi}\affiliation{\CNS}
\author{N.~Kobayashi}\affiliation{\Titech}
\author{T.~Kobayashi}\affiliation{\TOHOKU}
\author{Y.~Kondo}\affiliation{\Titech}
\author{Z.~Korkulu}\affiliation{\RIKEN}
\author{S.~Koyama}\affiliation{\TOKYO}
\author{V.~Lapoux}\affiliation{\CEA}
\author{Y.~Maeda}\affiliation{\MIYAZAKI}
\author{T.~Motobayashi}\affiliation{\RIKEN}
\author{T.~Miyazaki}\affiliation{\TOKYO}
\author{T.~Nakamura}\affiliation{\Titech}
\author{N.~Nakatsuka}\affiliation{\KYOTO}
\author{Y.~Nishio}\affiliation{\KYUSHU}
\author{A.~Obertelli}\affiliation{\CEA}\affiliation{\TUDA}
\author{A.~Ohkura}\affiliation{\KYUSHU}
\author{S.~Ota}\affiliation{\CNS}
\author{H.~Otsu}\affiliation{\RIKEN}
\author{T.~Ozaki}\affiliation{\Titech}
\author{V.~Panin}\affiliation{\RIKEN}\affiliation{\CEA}
\author{S.~Paschalis}\affiliation{\TUDA}
\author{E.C.~Pollacco}\affiliation{\CEA}
\author{S.~Reichert}\affiliation{\TUM}
\author{J.-Y.~Rousse}\affiliation{\CEAiv}
\author{A.T.~Saito}\affiliation{\Titech}
\author{S.~Sakaguchi}\affiliation{\KYUSHU}
\author{M.~Sako}\affiliation{\RIKEN}
\author{C.~Santamaria}\affiliation{\CEA}
\author{M.~Sasano}\affiliation{\RIKEN}
\author{H.~Sato}\affiliation{\RIKEN}
\author{M.~Shikata}\affiliation{\Titech}
\author{Y.~Shimizu}\affiliation{\RIKEN}
\author{Y.~Shindo}\affiliation{\KYUSHU}
\author{L.~Stuhl}\affiliation{\RIKEN}
\author{T.~Sumikama}\affiliation{\RIKEN}
\author{Y.L.~Sun}\affiliation{\CEA}\affiliation{\TUDA}
\author{M.~Tabata}\affiliation{\KYUSHU}
\author{Y.~Togano}\affiliation{\Titech}
\author{J.~Tsubota}\affiliation{\Titech}
\author{T.~Uesaka}\affiliation{\RIKEN}
\author{Z.H.~Yang}\affiliation{\RIKEN}
\author{J.~Yasuda}\affiliation{\KYUSHU}
\author{K.~Yoneda}\affiliation{\RIKEN}
\author{J.~Zenihiro}\affiliation{\RIKEN}
\date{\today}%
\begin{abstract}

 The structure and decay of the most neutron-rich beryllium isotope, $^{16}$Be, has been investigated following proton knockout from a high-energy $^{17}$B beam.
 \comment{Two relatively narrow resonances were observed for the first time, with energies of $0.84(3)$ and $2.15(5)$~MeV above the two-neutron decay threshold and widths of $0.32(8)$ and $0.95(15)$~MeV respectively.
 These were assigned to be the ground ($J^{\pi}=0^+$) and first excited ($2^+$) state, with $E_x=1.31(6)$~MeV.}
 The mass excess of $^{16}$Be was thus deduced to be $56.93(13)$~MeV, some $0.5$~MeV more bound than the only previous measurement.
 Both states were observed to decay by direct two-neutron emission.
 Calculations incorporating the evolution of the wavefunction during the decay as a genuine three-body process reproduced the principal characteristics of the neutron-neutron energy spectra for both levels, indicating that the ground state exhibits a strong spatially compact dineutron component, while the 2$^+$ level presents a far more diffuse neutron-neutron distribution.

\end{abstract}
\pacs{
21.10.Dr, 
25.40.-h, 
29.30.Hs, 
27.20.+n  
}
\maketitle


 \comment{The structure of nuclei lying far from beta stability represents a rich testing ground for our understanding of nuclear structure and other quantum phenomena owing to the large imbalance in the neutron-to-proton ratio. The light neutron-rich nuclei are of particular interest in this context as the dripline and beyond is experimentally accessible and they are amenable to being described by a wide range of theoretical approaches \cite{EdgeofStability}, ranging from the phenomenological shell-model \cite{ShellModel} to more ab initio approaches \cite{abinitio} as well as those incorporating explicitly the continuum \cite{GSMSpringer}. 
 Few-body effects, including correlations, play a central role in the structure of the most neutron-rich systems, as is most apparent in two-neutron halo nuclei whereby the halo neutron correlations are critical to the binding of the system. In more general terms, such systems have the potential to offer insight into the physics of open quantum systems \cite{MichelOpen}.}

 In the case of systems that lie two neutrons beyond the dripline, such as the subject of the present work $^{16}$Be \cite{Bau03}, two-neutron correlations play a crucial role and can, in principle, be probed through the system's decay into a core ($^{14}$Be) and two neutrons.
 Experimentally, the investigation of two-neutron decay is challenging and historically much effort has been given over to the exploration of two-proton decay \cite{Pfut12}.
 \comment{In that case}, however, the initial-state correlations are strongly perturbed by the effects of the Coulomb barrier and the Coulomb component of the proton-proton final-state interaction (FSI).

 In the last decade, advances in neutron detection techniques and the ability to produce sufficiently intense near-dripline beams have enabled a series of investigations to be made on two-neutron unbound systems \cite{Aksy08,Joha10,Kohl13,Spyr12,Lund12,Caes13,Kond16} and excited two-neutron continuum states \cite{Marq01,Aksy13,Hoff11,Jone15,Smith16,Reve18,Laur19}.
 The decay of excited states of $^{8}$He, $^{14}$Be and $^{20,24}$O, as well as the ground-state decay of $^{10}$He, exhibit signatures of sequential decay through resonances in the $A-1$ systems. In all cases where the $n$-$n$ observables were explored, an enhancement of the cross-section at low relative $n$-$n$ energies and/or angles was observed. Importantly, the interpretation of all of the measurements have employed rather simplified approaches, ignoring, for example, the initial state correlations or the effects of the FSI as the system decays.

 In the case of $^{16}$Be, which has been the subject of a single previous study \cite{Spyr12}, the relatively strong enhancement observed for low $n$-$n$ relative energies \comment{and correspondingly small opening angles} was interpreted as a signature of ``dineutron decay''.
 Specifically, comparison was made with two-body decay into $^{14}$Be and a quasibound dineutron, followed by the decay of the latter as described by the $n$-$n$ $s$-wave scattering length, and with three-body phase-space decay (no FSI between any of the decay products). The overly simplified character of this comparison was noted by Ref.~\cite{Marq12b}, \comment{including importantly the lack of consideration of the $n$-$n$ FSI in the three-body decay.}

 In this Letter we report on a new investigation of $^{16}$Be with greatly enhanced statistics, better resolution and superior acceptances. As a result, the ground and first excited states of $^{16}$Be have been clearly identified for the first time, allowing for an unambiguous and relatively precise determination \comment{of the mass of $^{16}$Be}. Furthermore, both states are clearly seen to decay by direct two-neutron emission. Comparison is made with results of three-body modeling of $^{16}$Be and its decay where, importantly, the time evolution of the initial-state wavefunction is taken into account.  
   

 The experiment was performed at the Radioactive Isotope Beam Factory 
 of the RIKEN Nishina Center.
 The secondary beam of $^{17}$B (with $E\sim277$~MeV/nucleon and $I\sim1.4\times10^4$~pps) 
 was produced by fragmentation of a 
 $^{48}$Ca primary beam on a thick Be target and prepared using the BigRIPS fragment separator \cite{Kubo03}. The beam particles were tracked event-by-event using two drift chambers onto the 15~cm thick liquid hydrogen target of MINOS \cite{MINOS}. The trajectories of the two protons from the $(p,2p)$ reaction were determined using the Time Projection Chamber 
 of MINOS, which permitted the reconstruction of the reaction vertex with a resolution ({\sc fwhm}) of $\sim5$~mm \cite{Belen_PhD}.
 The use of a thick target combined with the determination of the vertex allowed for a significant enhancement in the luminosity whilst maintaining a good invariant-mass resolution.

\begin{figure}
 \begin{center}
 \hspace*{-2mm}\includegraphics[width=.50\textwidth]{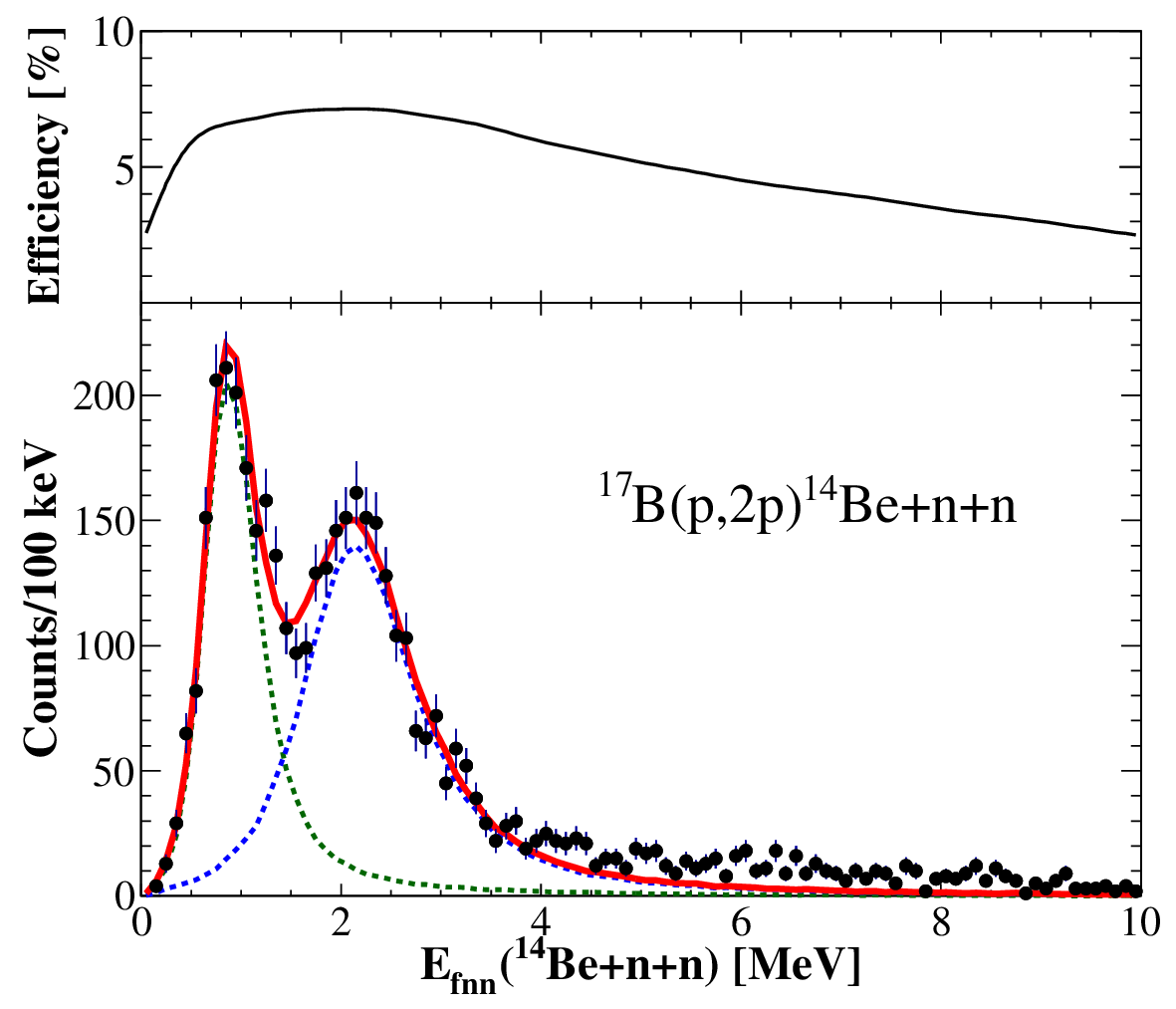}
 \caption{(Color online) Relative-energy spectrum of $^{14}\mbox{Be}+n+n$ events ($E_{fnn}$) following the $^{17}$B$(p,2p)$ reaction. The red line represents the best fit, \comment{incorporating all experimental effects,} up to 4~MeV ($\chi^2/{\rm{ndf}}=1.3$) including $^{16}$Be resonances at $0.84$ and $2.15$~MeV (dotted lines). The upper panel displays the overall detection efficiency, including the effects of the neutron cross-talk rejection filter.} \label{f:Efit}
 \end{center}
\end{figure}
 
 The forward going beam-velocity charged fragments and neutrons were detected using the SAMURAI spectrometer \cite{Koba13,Shimizu} and the neutron array NEBULA \cite{Nak16,Kondo20}. Significant care was taken to eliminate cross-talk events (neutrons detected more than once in the array) offline, with the percentage of such events being reduced to less than $\sim4$\% \cite{Belen_PhD}.
 The energy of the unbound $^{16}$Be was reconstructed from the relative energy of the $^{14}\mbox{Be}+n+n$ decay products ($E_{fnn}$) as the invariant mass of the system minus the masses of its constituents (Fig.~\ref{f:Efit}). As $^{14}$Be has no bound excited states, the energy so determined reflects directly the energy above the $2n$ emission threshold.
 The resolution in relative energy varied as $\sim0.5\sqrt{E_{fnn}}$~MeV \cite{Belen_PhD} whilst the efficiency varied smoothly with relative energy and was around 5\% in the range of interest (Fig.~\ref{f:Efit}).
 Further details of the setup, simulations and analysis techniques, including the neutron cross-talk rejection filter and associated verifications, may be found in Refs.~\cite{Kond16,Belen_PhD,Nak16,Kondo20,Corsi_PLB,Kubota_PRL}.


 The energy spectrum of $^{14}\mbox{Be}+n+n$ events shown in Fig.~\ref{f:Efit} is dominated by two strongly populated resonance-like structures in the region below 4~MeV. The spectrum was fit in this range employing two Breit-Wigner lineshapes with energy-dependent widths \cite{BW} as inputs for a complete simulation of the setup, secondary beam characteristics and reaction process \cite{Belen_PhD}.
 As a result, $^{16}$Be resonance energies of $E=0.84(3)$ and $2.15(5)$~MeV, and widths of respectively $\Gamma=0.32(8)$ and $0.95(15)$~MeV, were obtained. Whilst the slight excess of counts, with respect to the best fit, for energies above $\sim4$~MeV might suggest a very small contribution from the non-resonant continuum and/or very broad, weakly populated higher-lying structures, their inclusion makes no discernible changes to the results for the resonances \cite{Belen_PhD}.  
  
 \begin{figure}
 \begin{center}
 \includegraphics[width=.42\textwidth]{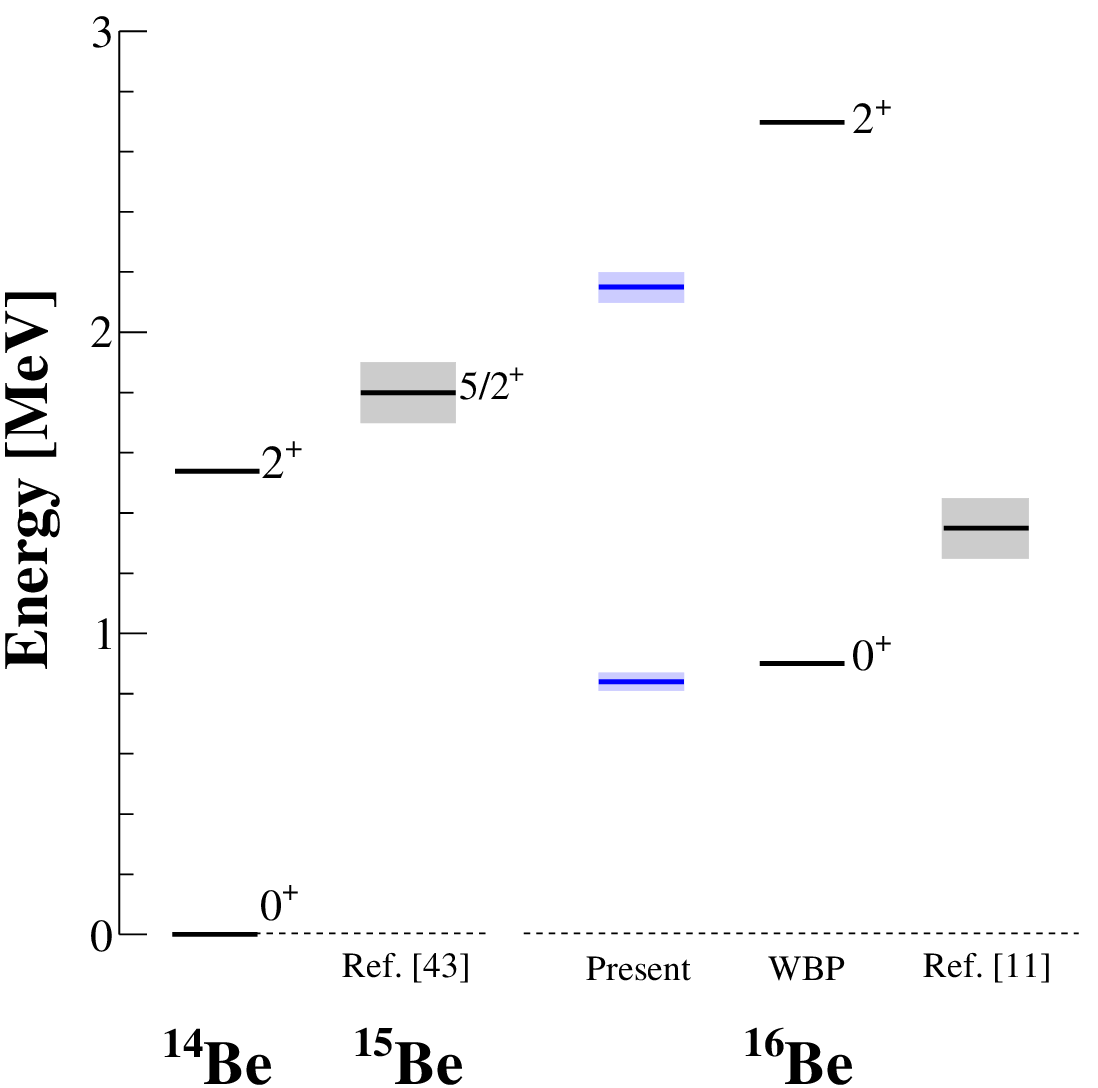}
 \caption{(Color online) Low-lying level schemes of the most neutron-rich beryllium isotopes. The shaded bands represent the error bars of the measured energies. The three columns on the right correspond to $^{16}$Be present results, shell-model calculations (WBP) \cite{Spyr12}, and the result of Ref.~\cite{Spyr12}.} \label{f:levels}
 \end{center}
\end{figure}
 
 \comment{Combining the energy of the lowest lying resonance (or $S_{2n}$) with the mass of $^{14}$Be \cite{AME20} allows the mass excess of $^{16}$Be to be determined as $56.93(13)$~MeV, where the uncertainty is dominated by that of $^{14}$Be. This is $0.52$~MeV more bound than the mass excess of $57.45(17)$~MeV derived from the result of Ref.~\cite{Spyr12}, and $0.77$~MeV more bound than the mass-surface extrapolation of $57.68(50)$~MeV \cite{AME03}\footnote{More recent mass evaluations list the experimental result of Ref.~\cite{Spyr12} and no mass-surface extrapolations are made.}.
 The differences in the present results with respect to the study of Spyrou \etal\ \cite{Spyr12}, which also employed proton removal from $^{17}$B but only identified a single broader ($\Gamma=0.8^{+0.1}_{-0.2}$~MeV) structure at $1.35(10)$~MeV,} arise from the much enhanced luminosity of the present experiment coupled, importantly, with a superior detection efficiency at relative energies away from threshold and a better resolution.
 The present result, when combined with others \cite{JLL09,Yang21,Spyr10,Gaud12} not available at the time of the compilation of Ref.~\cite{AME03}, should allow for more reliable mass-surface extrapolations to be made for neighboring nuclei, including $^{17,18}$Be, and may thus provide a guide to their possible existence as identifiable resonances in the $3n$ and $4n$ continua.

 As an even-even nucleus, the low-lying level structure of $^{16}$Be will comprise a $J^\pi=0^+$ ground state and very probably a $2^+$ first excited state, which we associate with the two resonances observed here. As such the excitation energy of the 
$2^+$ level is $1.31(6)$~MeV, \comment{making it the lowest lying $2^+$ state in the beryllium isotopic chain \cite{NNDC,Sugi07}.}
 Shell-model calculations performed in the $s$-$p$-$sd$-$pf$ model space with the WBP Hamiltonian \cite{Spyr12} predict the ground state to be unbound to $2n$ emission by $0.9$~MeV and the first excited state to be a $2^+$ level at $2.7$~MeV ($E_x=1.9$~MeV), in good agreement with the present observations (Fig.~\ref{f:levels}). No results are yet available from ab initio approaches or the Gamow Shell Model.
 \comment{However, the three-body model employed here and detailed below is crafted explicitly to explore continuum states and predicts, using the present energy for the ground state as an input, the 2$^+$ level at $E_x=1.3$~MeV.}

 Turning to the two-neutron decay of $^{16}$Be, Fig.~\ref{f:levels} displays the level schemes of $^{14-16}$Be\footnote{Two other excited states in $^{14}$Be lie above 3~MeV \cite{Aksy13}.}. Levels in $^{15}$Be are of importance in terms of whether energetically the states in $^{16}$Be can decay sequentially via $^{15}$Be and also in defining the $^{14}$Be-$n$ interaction for the three-body modeling of $^{16}$Be.
 Although the shell model \cite{Lovell17} and systematics of the $N=11$ isotones \cite{Suso20} predict the lowest-lying levels in $^{15}$Be to be $J^\pi=5/2^+,1/2^+,3/2^+$, only one level, believed to be the $5/2^+$ and lying $1.8(1)$~MeV above the $1n$ threshold with a width of 0.58(20)~MeV, has been observed in a neutron transfer study with a $^{14}$Be beam \cite{Snyd13}.
 A series of studies using nucleon removal and fragmentation type reactions \cite{Spyr11,Belen_PhD,Belen_PRC}, including a search for the $3n$ decay of levels via the unbound $2^+$ state of $^{14}$Be \cite{Kuch15}, have not identified any other state. As such, energetically the $^{16}$Be ground state must decay by direct two-neutron emission to the $^{14}$Be ground state, whereas the $2^+$ level may decay sequentially via the known $^{15}$Be resonance. It should be noted that decay via any levels in $^{15}$Be which in turn decay via two-neutron unbound $^{14}$Be$(2^+)$ to $^{12}$Be will not be observed in the present channel. 

\begin{figure}
 \begin{center} \def\wY{36mm}\def\wX{42mm}
 \hspace*{-1mm}\psfig{file=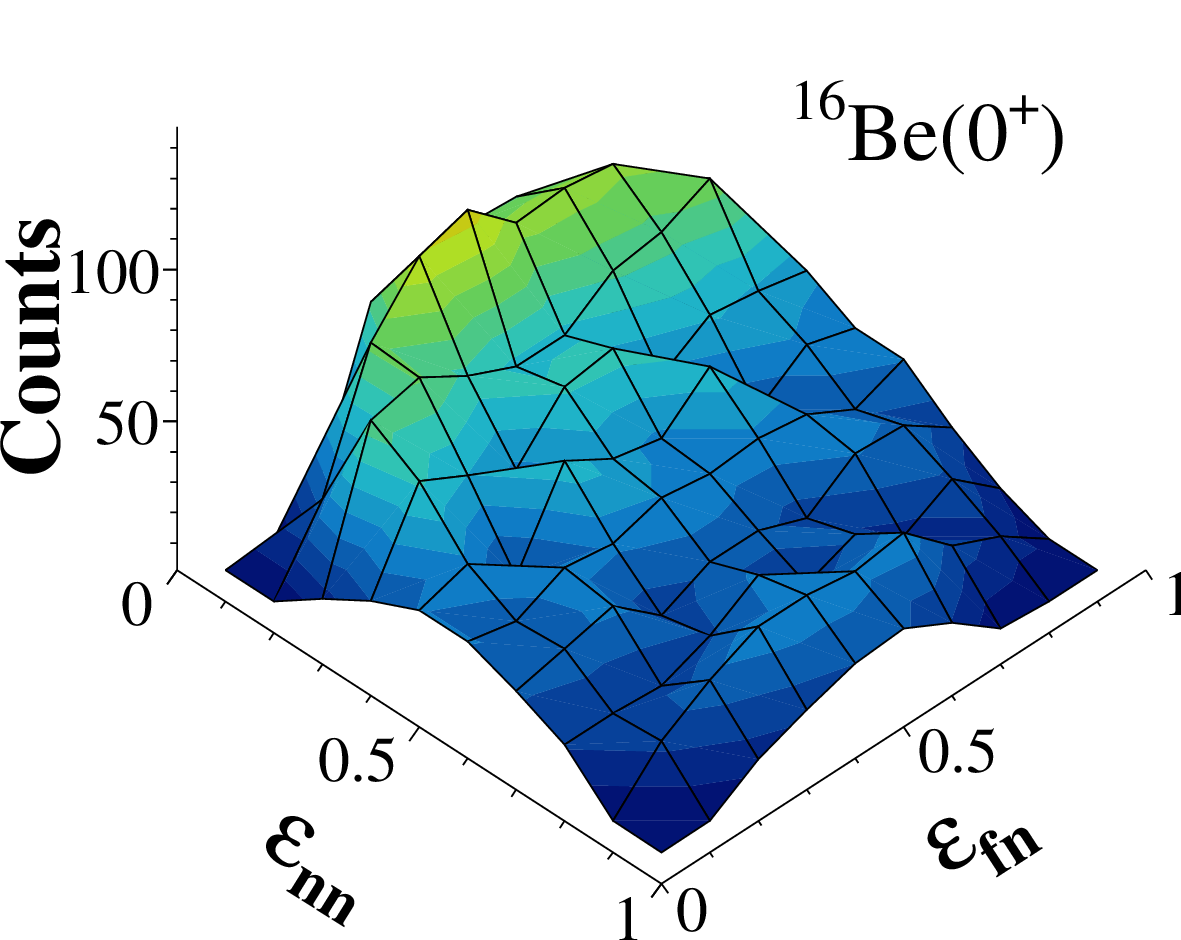,height=\wY,clip=}\psfig
                    {file=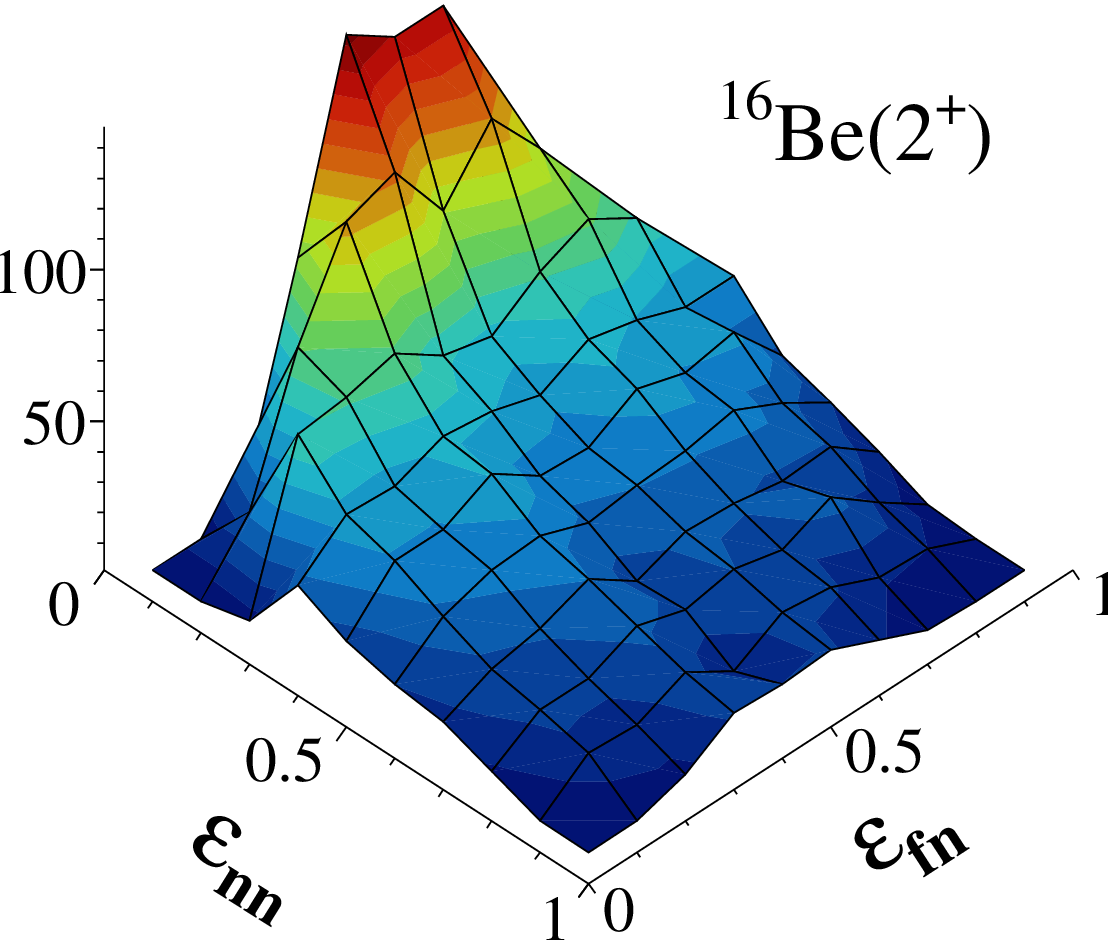,height=\wY,clip=}\\[4mm]
 \psfig{file=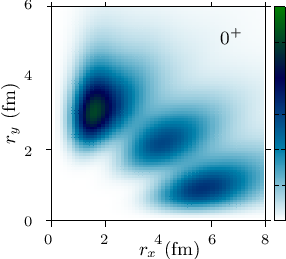,width=\wX}\hfill\psfig{file=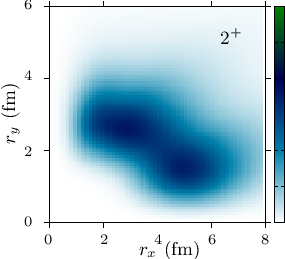,width=\wX}
 \caption{(Color online) Top: experimental Dalitz plots of the normalized energies $^{14}$Be-$n$ ($\varepsilon_{fn}$) vs $n$-$n$ ($\varepsilon_{nn}$) for the decay of the two observed states in $^{16}$Be, selected by gates in $E_{fnn}$ between 0--1.2 and 1.7--3.0~MeV (see Fig.~\ref{f:Efit}). Bottom: theoretical spatial probability distributions $P(r_x,r_y)$ for each state in terms of the distances $^{14}$Be-$nn$ ($r_y$) vs $n$-$n$ ($r_x$), with the color scale ranging from 0 to 0.12 fm$^{-2}$.} \label{f:Edal}
 \end{center}
\end{figure}

 Experimentally, the characteristics of the decay have been investigated, as shown in Fig.~\ref{f:Edal}, using Dalitz plots of the normalized relative energies $\varepsilon_{ij}=E_{ij}/E_{fnn}$ (equivalent to the normalized invariant masses employed in Refs.~\cite{Marq01,Reve18,Laur19}). In the absence of any interaction, three-body phase-space decay leads to a uniform distribution within the kinematical boundary of the plot. A clear enhancement, however, is visible at low $\varepsilon_{nn}$ for both $^{16}$Be states.
 No evidence is seen for sequential decay via states in $^{15}$Be, which would manifest itself as bands in $\varepsilon_{fn}$ corresponding to $^{14}$Be-$n$ FSI \cite{Marq01,Reve18}. It is clear, therefore, that both the ground and $2^+$ levels of $^{16}$Be decay via direct two-neutron emission. 

 In order to explore what may be deduced regarding the structure of $^{16}$Be from the observed two-neutron decay, the natural avenue is comparison with three-body ($^{14}\mbox{Be}+n+n$) modeling, incorporating a realistic description of the decay. The approach used here to construct the $^{16}$Be wavefunction in the continuum is described in detail in Refs.~\cite{Casal18,Casal19} and is based on the standard Jacobi coordinates for three-body systems using the hyperspherical formulation \cite{Zhukov93}. As such, the configurations compatible with the total $J^\pi$ of the system are labeled $\{K,\ell_x,\ell_y,\ell,S_x\}$ where: $\ell_x,\ell_y$ denote the relative orbital angular momentum between the neutrons and between the neutron pair and the core, respectively; $\ell$ is the total orbital angular momentum; $S_x$ is the total spin; and $K$ is the so-called hypermomentum that defines the effective barriers in the hyperradial coupled-channels system.
 
 The principal feature of the method is the definition of a resonance operator, the eigenvalues of which describe localized continuum structures as a combination of discretized continuum states of different energy. This allowed the lowest $0^+$ and $2^+$ states in the $^{14}\mbox{Be}+n+n$ continuum to be identified.
 Calculations were carried out using the Gogny-Pires-Torreil $n$-$n$ potential and a core-$n$ potential fixed by the energy of the $d$-wave resonance of $^{15}$Be, supplemented by a two-parameter Gaussian three-body force in order to fix the energy of the $^{16}$Be $0^+$ state at $0.85$~MeV above the $2n$ threshold (in Ref.~\cite{Casal19} the energy from Spyrou \etal~\cite{Spyr12} was employed).
 The two-neutron probability densities determined for both states are shown in Fig.~\ref{f:Edal} as a function of the distances $^{14}$Be-$nn$ ($r_y$) and $n$-$n$ ($r_x$). In the case of the $0^+$ state, the density distribution, which is characterized by the three-lobe structure of an essentially pure $d$-wave configuration, exhibits a strong spatially compact dineutron admixture. In contrast, the $2^+$ level presents a quite diffuse neutron-neutron spatial distribution.
 In the case of the only other three-body calculation of the structure of $^{16}$Be \cite{Lovell17}, the ground state was also found to be dominated by a compact dineutron configuration, but no predictions were made for the energy or configuration of the 2$^+$ level.

\comment{As indicated above, the energy of the $2^+$ resonance was computed to be $2.15$~MeV ($E_x=1.30$~MeV), in very good agreement with the excited state observed here. The resonance widths were also evaluated and values of $\Gamma(0^+)=0.10$ and $\Gamma(2^+)=0.42$~MeV were found, both smaller than experiment. The model at present, however, treats the $^{14}$Be core as spherical and inert \cite{Casal18} and the inclusion of deformed core excited states may increase the predicted widths \cite{Casal19}.
 It may be noted that in Ref.~\cite{Lovell17} a width of 0.17~MeV was calculated for the ground state\footnote{An almost identical width of 0.16~MeV was predicted by the present model \cite{Casal18} when employing the ground state energy of Ref.~\cite{Spyr12}.}, while a value of 0.42~MeV was estimated in a simplified picture of $2n$ cluster decay \cite{HTF16Be}. In both cases, however, the input in terms of the energy of the ground state was the higher value of the earlier study of Ref.~\cite{Spyr12} (Fig.~\ref{f:levels}).}

\begin{figure}
 \begin{center}
 \includegraphics[width=.44\textwidth]{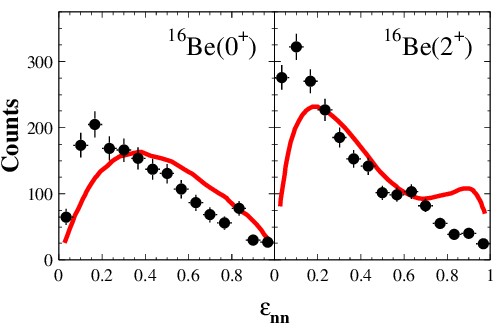}
 \caption{(Color online) Normalized $n$-$n$ energy distributions for the $0^+$ and $2^+$ levels compared to the predictions of the three-body model incorporating a realistic description of the decay (red lines, see text).} \label{f:Enn}
 \end{center}
\end{figure}

 In order to investigate the relationship between the measurements and the predicted structure of the $^{16}$Be states (Fig.~\ref{f:Edal}), the wavefunctions, which are not stationary states, were used to find the solution of an inhomogeneous equation where the source term takes into account all interactions ($n$-$n$, core-$n$ and three-body) \cite{Casalnew}. This provides the correlations between the three bodies produced asymptotically, in the spirit of the calculations of Ref.~\cite{Gri09prc}.
 \comment{Fig.~\ref{f:Enn} displays the results for the $n$-$n$ energy distributions ($\varepsilon_{nn}$)\footnote{The $n$-$n$ relative energy distribution (combining the relative momentum and angle information) provides for the most complete probe of the corresponding correlations.}}, after filtering through the simulations which account for the experimental effects.
 The experimental $\varepsilon_{nn}$ spectra were obtained by fitting the $E_{fnn}$ spectra with the two resonances (Fig.~\ref{f:Efit}) for each bin in $\varepsilon_{nn}$.

 The comparison shows that the calculations capture the overall trend of the measurements, which exhibit enhancements for both states at small $n$-$n$ relative energy. In particular, the calculations for the decay of the $2^+$ level exhibit a more pronounced enhancement at small $\varepsilon_{nn}$, although the initial state presents a rather diffuse $n$-$n$ spatial distribution as compared to the $0^+$ level (Fig.~\ref{f:Edal}). 
 This reflects, in part, the influence of the corresponding initial state $n$-$n$ momentum distributions which, in simple terms, will favor small relative momenta for the $2^+$ level and the contrary for the more compact dineutron-like configuration that dominates the $0^+$ state.

 Within the calculations, the more detailed form of the relative-energy distributions is governed by the interference between the lowest-$K$ configurations, with weights that shift from the dominant $K=4$ terms in the initial state to mostly $K=0,2$ asymptotically. This leads to an increase of the relative $s$-wave components since they are subject to smaller centrifugal barriers, as discussed in a more general context in the time-dependent method of Ref.~\cite{Wang21}.
 The calculations for the decay of the $2^+$ state deviate from experiment above $\varepsilon_{nn}\sim0.7$, with theory exhibiting a minimum followed by a rise at high $\varepsilon_{nn}$. This behavior is attributed to interference between the $K=2$ and $4$ components of the wavefunction.
 It remains to be determined whether more complete calculations, including $^{14}$Be core excited states and the inclusion in the core-$n$ potential of angular momentum channels beyond that defined by the $^{15}$Be $5/2^+$ resonance (namely the $1/2^+$ and $3/2^+$ states when experimentally located) could change the $K$ and $\ell$ admixtures and further improve the agreement with experiment.

 In more general terms the present work underlines the need to go beyond naive descriptions of two-neutron decay, such as that invoked in earlier studies \cite{Kohl13,Smith16}, including that of $^{16}$Be \cite{Spyr12}, and employ realistic wavefunctions that are properly time evolved to describe the three-body decay.  
 In a similar vein, techniques that employ only the effects of the $s$-wave $n$-$n$ FSI \cite{Led82} to derive average $n$-$n$ separations in the initial states of these systems \cite{Marq01,Reve18,Laur19,Marq00} also suffer from major deficiencies. 
 Indeed, in such an approach our measured $n$-$n$ energy distributions would be interpreted in terms of a rather compact $n$-$n$ spatial configuration for the $2^+$ state and a significantly more diffuse one for the $0^+$ ground state \cite{Belen_PhD}, in contrast with the microscopic predictions.


 In summary, the structure and decay of the heaviest known beryllium isotope, $^{16}$Be, has been investigated following proton knockout from a high-energy $^{17}$B beam. The study benefited from the enhanced luminosity offered by an intense secondary $^{17}$B beam coupled with a thick liquid hydrogen target, incorporating vertex detection, and a large acceptance setup.
 The ground ($0^+$) and first excited ($2^+$) states were observed for the first time as relatively narrow resonances, at $0.84(3)$ and $2.15(5)$~MeV above the $2n$ decay threshold \comment{with widths of $0.32(8)$ and $0.95(15)$~MeV respectively.}
 A ground-state mass excess of $56.93(13)$~MeV was thus determined, some $0.5$~MeV more bound than the only previous measurement \cite{Spyr12}.
 The excitation energy of the $2^+$ level, $1.31(6)$~MeV, is in good accord with WBP interaction shell-model calculations \cite{Spyr12} as well as the three-body modeling of $^{16}$Be presented here.  
 Both states were found to decay via direct two-neutron emission to the $^{14}$Be ground state, despite sequential decay via $^{15}$Be being energetically allowed in the case of the $2^+$ level.

 Realistic three-body modeling incorporating, importantly, the asymptotic properties after the time evolution of the initial resonance wavefunction, was employed to explore the two-neutron decay. The calculations were seen to reproduce the principal features of the neutron-neutron energy as a result of a genuine three-body decay for both the ground and excited states of $^{16}$Be.
 In the case of the former, the wavefunction exhibited a strong compact dineutron admixture, while the latter presented a much more diffuse neutron-neutron spatial distribution. In more general terms, the approach presented here demonstrates the importance of realistic modeling including the decay process itself in order to investigate two-neutron decay.

\header{Acknowledgments}
The authors wish to extend their thanks to the accelerator staff of the RIKEN Nishina Center for their efforts in delivering the intense $^{48}$Ca beam. J.G., F.M.M.\ and N.A.O.\ acknowledge partial support from the Franco-Japanese LIA-International Associated Laboratory for Nuclear Structure Problems as well as the French ANR-14-CE33-0022-02 EXPAND.
 J.C.\ acknowledges financial support by the European Union's Horizon 2020 research and innovation program under the Marie Sklodowska-Curie grant agreement No.\ 101023609, by MCIN/AEI/10.13039/501100011033 under I+D+i project No.\ PID2020-114687GB-I00, and by the Consejer\'{\i}a de Econom\'{\i}a, Conocimiento, Empresas y Universidad, Junta de Andaluc\'{\i}a (Spain) and ``ERDF-A Way of Making Europe'' under PAIDI 2020 project No.\ P20\_01247. J.G.C.\ is supported by MCIN/AEI/10.13039/501100011033 under I+D+i project No.\ PID2021-123879OB-C21.
 This work was also supported in part by JSPS KAKENHI Grant No.~24740154 and 16H02179, MEXT KAKENHI Grant No.~24105005 and 18H05404, the WCU (R32-2008-000-10155-0), the GPF (NRF-2011-0006492) programs of the NRF Korea, the HIC for FAIR, the CUSTIPEN (China-US Theory Institute for Physics with Exotic Nuclei) funded by the US Department of Energy, Office of Science under grant number DE-SC0009971, the Office of Nuclear Physics under Awards No.~DE-SC0013365 (MSU) and No.~DE-SC0018083 (NUCLEI SciDAC-4 Collaboration), and the European Research Council through the ERC Starting Grant No.\ MINOS-258567. 


\end{document}